\documentclass[11pt,epsf] {article}
\usepackage{epsfig,wrapfig,color,sidecap}
\usepackage{setspace}
\usepackage{index}
\usepackage{subfigure}
\usepackage[compact]{titlesec} 
\titlespacing{\section}{0pt}{2ex}{1ex}
 \titlespacing{\subsection}{0pt}{1ex}{0ex} 
 \titlespacing{\subsubsection}{0pt}{0.5ex}{0ex}
\usepackage[top=2.5cm, bottom=2.5cm, left=2.cm, right=2.cm]{geometry}
\usepackage{color}
\usepackage{multicol}
\newcommand{\bfig}{\begin{figure}}
\newcommand{\efig}{\end{figure}}
\newcommand{\btab}{\begin{table}}
\newcommand{\etab}{\end{table}}
\newcommand{\btabwide}{\begin{table*}}
\newcommand{\etabwide}{\end{table*}}

\definecolor{blue}{rgb}{0.,0.,0.99}
\definecolor{black}{rgb}{1.,1.,1.}
\usepackage{paralist}
\makeindex

\begin{document}

\title{ Did Edwin Hubble plagiarize?}
\author{ Giora Shaviv \\
Department of Physics, Technion\\
Israel Institute of Technology, Haifa, Israel, 32,000}

\maketitle

\abstract{Recently Block published an astro-ph{\footnote{http://arxiv.org/abs/1106.3928 (2011).}} insinuating that Lemaitre discovery paper of the Expanding Universe was censored   prior to its translation into English and publication in the Monthly Notices of the Royal Astronomical Society. Consequently, Lemaitre's credit for the discovery of the velocity-distance correlation was not recognized. We examine here the chain of events leading to the discovery of the 'Hubble law'. 
Our summary: (a) Lemaitre found a theoretical linear correlation between velocity and distance. (b) Lemaitre assumed the existence of a linear relation between velocity and distance and calculated the coefficient. (c) Hubble took the data plotted it and demonstrated that a linear relation represents the observed data.  (d) Hubble never believed in Lemaitre's solution, namely in an expanding universe. Consequently, Hubble never cited Lemaitre.   We conclude that the charge that Lemaitre's  paper was censored or ignored let alone plagiarized by Hubble, is not founded , and explain why Lemaitre's earlier theoretical discovery and derived 'Hubble constant' was not cited or  recognized, by Hubble as well as by many other leading researchers. 

\section{~The Discovery of the Expanding Universe - theory}
As soon as  Einstein published in 1915, in a series of papers,  his General Theory of Relativity, attempts to apply the new theory to the entire universe were made by Einstein{\footnote { Einstein, A., Berl. Ber., 1917, p 142., Ann. d. Phys. {\bf 55}, 241, (1918)}} himself and by  de Sitter (1872-1934){\footnote {de Sitter, W., Proc. Akad. Wetesch. Amsterdam,  {\bf 19}, 1217, (1917), Ibid. {\bf  20}, 229, (1917)}}. Both  assumed an empty  universe without mass, more accurately, they ignored the pressure inside the universe.  Einstein searched for a static solution  while de Sitter allowed for a dynamic one. It would  be found later  that technically, de Sitter's solutions are also static, but this was due to the assumption that his models assumed an empty universe, namely in his approximation the mass did not affect the results.  

Then de Sitter calculated the gravitational field produced by a hollow spherical shell, whose total mass is M, and whose inner and outer radii are $R_1$ and $R_2$, respectively. de Sitter compared the units of space and time inside and outside the hollow sphere and found them to change in a way that {\it cannot be verified by observation of the motions of bodies inside the shell relatively to each other.} de Sitter then proceeded and argued that the difference in the unit of time can, however, be verified by observing light which emanates from a point within the mass, or outside the shell. According to de Sitter's solution, {\it light coming from distances must appear displaces towards the violet as compared with light from a source inside the shell },  an effect which became known as the 'de Sitter effect'{\footnote{When de Sitter described Lemaitre's solution in 1930 (BAN, {\bf 5}, 211, 1930) he named the effect {\it the expanding universe}}}. In other words, de Sitter's effect is a result of space geometry affected by gravitation where the geometrically induced red-shift would appear as an apparent Doppler induced velocity. According to de Sitter, the result emerges from the fact that we always measure distance and time in certain combinations and never say, distance alone. Also, it should be recalled that at this moment in time de Sitter looked for a steady state solution. In the second paper de Sitter derived a formula for the {\it velocities due to inertia} and noticed that there is no preference of sign. He then compared three nebula{\footnote{The nebulae are: Andromeda, NGC 1068 and NGC 4594 with velocities: $-311km/sec$,  $925km/sec$ and  $1185km/sec$, respectively.}} and argued that the speeds are very large and show no preference in sign. de Sitter then took the mean of the three(!) velocities to find $600km/sec$ and assumed an average distance of $10^5pc$ to obtain a radius of curvature of $3\times 10^{11}AU$. He however, admitted that this result has no practical value, but if you want you can say that the first measurement of the distance-velocity relation by de Sitter yielded in 1917, $6000km/sec~per~Mpc$.

de Sitter's solution was found before  the Great Debate on the nature of the nebulae was resolved, and de Sitter did not know at the time of the writing that  extragalactic objects   exist. Consequently, de Sitter{\footnote{de Sitter, W.,  MNRAS, {\bf 77}, 155, (1916), ibid., {\bf 78}, 3, (1917)}} could only speak on  distant stars, and  estimated that the shift observed in faint  and hence distant stars, would be  equivalent to less than $1/3 ~km/sec$. Happy with this result, de Sitter wrote that:  {\it It is well known that stars of all spectra actually show a small systematic displacement towards the red},  not the displacement to the violet as he predicted.   Einstein did not like the solution de Sitter found and criticized it.  Assuming that the shift of stars to the blue does not exceed $1km/sec$, de Sitter calculated the total mass within the radius of curvature of the Universe.

It was 
Lanczos (1893-1974) who demonstrated{\footnote {Lanczos, C., Phys. Zeit., {\bf 23}, 539, (1922)}},  by a simple change of the coordinates, that the de Sitter static solution can be interpreted as an expansion of the universe and the de Sitter predicted red-shift is a genuine recession velocity. However, the direct solutions  for the expanding universe (without manipulations with the coordinates) were found by Alexander Friedman (1888-1925){\footnote {Friedman, A.A., Zeit. fur Phys. {\bf 10}, 377, (1922), Ibid., {\bf 21}, 326, (1924)}} in 1924. Friedman's solutions are known today as the 'Friedman world models'. Einstein believed in 1923 that he found an error in the 1922 Friedman's paper because it did not agree with his expectations,  but Friedman demonstrated  that this was not the case. Consequently, Einstein withdrew his objection to the result. It was difficult to shake Einstein's belief in the static universe.  A summary of the static versus the non-static solutions to Einstein's  equations was published by Robertson in 1933{\footnote {Robertson, H.P., Rev. Mod. Phys.,{\bf 5}, 62, (1933)}}, when it was already clear that static solutions do not agree with the observations. 

In 1927 the Catholic Jesuit priest Abb${\acute{\rm e}}$  Lemaitre (1894-1966) discovered, being unaware of  Friedmann's, solutions to Einstein's equation which included mass and pressure and effectively rediscovered  that Einstein's  solution for the universe is  unstable, namely, they behave like a giant explosion. It is in this paper that Lemaitre derived the {\it apparent Doppler effect} which implied that {\it the receding velocities of extragalactic nebulae are a cosmical effect of the expansion of the universe} with velocity proportional to the distance.

In contrast to Friedman,  who was happy in deriving only the  theory, Lemaitre then turned to observation in an attempt to verify the prediction. To that goal he used a list of 42 nebulae and their velocities  which was given by Str\"omberg{\footnote{Str\"omberg, ApJ, {\bf 61}, 353, (1925)}}. As for the distance of the nebulae, Lemaitre followed Hubble{\footnote{Lemaitre wrote that the distances are from Hubble but gave no reference. We suspect that it is: Hubble, E., ApJ, {\bf 64} 321, (1926).}}, who has shown that  the extra-galactic nebulae have all the same absolute magnitude of -15.2. If so, the distance $r$, in parsecs, of an extra-galactic nebula with apparent magnitude $m$ is $log~r=0.2m+4.04$. Since the error is expected to rise with the distance, Lemaitre chose to assign a weight of $1/\sqrt{1+r^2}$ to each nebula at a distance $r$ where $r$ is expressed in Mpc. 

What Lemaitre did next was to correct the velocities for the speed of the sun and calculated the mean distance and the mean velocity and found a mean velocity of $600km/sec$ and mean distance of $0.95Mpc$ which imply, a coefficient of  $625km/sec~per~Mpc$. More accurately, if  you assume a linear relation between the velocity and distance then the coefficient is the one found by Lemaitre, but Lemaitre did not prove that the observations imply a linear relation. The theory does.   

It is interesting to learn what Lemaitre thought about the observational  data. In a  footnote (page 36) he stated that (my translation from french): {\it If one does not attribute weights to the observations, one finds $670 km/sec$ per $1.16\times 10^6$ parsecs or about $575 km/sec~per~Mpc$. Certain authors looked for a correlation between $v$ and $r$ and obtained only a very weak correlation.  The error in the distance determinations of the individual objects is of the order covered by the observation interval and the proper velocity of the nebulae is large ($300km/sec$   according to Str\"omberg). Accordingly, it seems that the negative results are neither for nor against the  relativistic interpretation of the Doppler effect. All what the inaccuracy in the observations allows us to do is to suppose that $v$ is proportional to $r$ and to try to avoid a systematic error in the determination of $v/r$. cf.  Lunmark } for the error estimate, which  Lemaitre did not provide.

I went to the   Universit$\grave{\rm e}$ Catholic the Louvain-la-Neave (UCL), where Lemaitre spent most of his professional life and he is immortalized in many ways  and checked in the library for all Lemaitre's papers published before 1930. There was only one paper, the transcript of Lemaitre's talk: {\it The Size of the Cosmos}, a talk delivered by Lemaitre at the Brussels scientific association on January 31, 1929{\footnote{Lemaitre, G., {\it La Grandear de l'Espace}, Conference faite le 31, Janvier 1929 a la Societe Scientificue de Bruxelles, Paris, Les Presses Universitaires de France. March 1929.}}. In the transcript, which is 23 page long, Lemaitre brings all the data which supported de Sitter's model and he did not mention at all his model from 1927. Lemaitre described the mean density of galaxies in the universe, the magnitude of galaxies, as well as  the fact that of the about 42 known galaxies with measured radial velocities all but three recede from us.  Lemaitre even mentions that there is a relation between the velocity of recession and the distance, but does not give  any number which connects the velocity of recession with the distance, namely the number he himself derived in 1927.  In other words, the paper contains many numbers but not any estimate of the Hubble constant.

The theoretical result found by Lemaitre was also found by Robertson in 1928{\footnote {Robertson, H.P., Phil. Mag., {\bf 5}, 835, (1928). Ibid.,  Phil. Mag., Supple.,  {\bf 5}, 385, (1928)}}  who,  so it seems,  did not know about   Lemaitre's 1927  paper. Robertson noted in his second  paper that he discovered Lemaitre's 1925 paper{\footnote {Lemaitre, G., J. Math. and Phys.,{\bf  4}, 188, (1925)}} only after the completion of his paper, but apparently he was unaware of  Lemaitre's important paper of 1927. 

In these early stages of the developments of the ideas, it was not clear what exactly happened $'$at the beginning$'$ of the universe. Was there a singularity, namely, the density and temperature tend to infinity, or the universe started expanding from a static state and if so what was the static state?   A l${\acute{\rm a}}$ Lemaitre, the  universe started $'$from some primeval matter$'$, the properties of which were ill defined.

\section{~The K-term}
As soon as the velocities of stars were measured, it was discovered that the sum of all stellar  velocities  relative to the sun does not vanish. So to force the vanishing of the mean an extra constant was artificially added. This constant was called the K-term or the K-effect. The non vanishing term  is the K-effect. Soon astronomers discovered that different classes of stars  have different K-terms. However, the stellar K-term is never larger than $10km/sec$.

\section{Discovering the recession of the nebulae}

In 1912 Slipher (1875-1969) reported on    first Doppler measurement of the radial velocity of the Andromeda Nebula ($-300km/sec$){\footnote{Slipher, V.M.,  LowOB, {\bf 2, 56, (1913)}}.  Slipher noticed that:  {\it The magnitude of this velocity, which is the greatest hitherto observed, raises the question whether the velocity-like displacement might not be due to some other cause, but I believe we have at present no other interpretation for it.}

The equipment Slipher used is amazing. Slipher attached his spectrograph to a 24-inch refractor. In some cases Slipher   applied exposure times as long as 40 hours. 
 
In 1917 Slipher {\footnote{Slipher, V.M., PAPhS, {\bf 56}, 403, (1917)}} published a list of 20 nebulae with their Doppler shift.  This discovery, which was the basis for over a decade of attempts to find a velocity distance correlations, is considered as one of the most important discoveries of the Lowel Observatory.  The confusing fact was that the 25 nebula were not isotropically distributed in the sky. Most of the positive velocities were close to one direction and the few of the nebula in the opposite direction had negative velocities. So it appeared a bit premature to claim the expansion of the nebulae system. This unisotropy, which eventually turned out to be accidental, was ignored towards the end of the nineteen twenties and the neglect allowed the discovery of the distance-redshift correlation, but it did not prevent attempts to find velocity-direction correlation.

  The fact that the nebulae show mostly a shift to the red  of the spectral lines was   known already to  Campbell{\footnote{ Campbell, W.W., {\it Note on Radial Velocities of nebulae}, AN, {\bf 188}, 346, (1911)}} as early as  1911 and to   Slipher{\footnote {Slipher, V.M., {\it Spectrographic Observations of Nebulae}, Popular Astronomy, {\bf 23}, 21, (1915)}} in 1915,  long before the location of the nebulae (inside or outside our Milky Way) was known. Slipher measured 14 spiral nebulae  and out of these,  two showed negative velocities  and three showed no velocity at all. All the rest showed positive velocity, namely, they recede from us. Slipher explained the results by noticing that the velocities of the nebulae were about 25 
times greater than the average stellar velocity{\footnote{Reynolds (Obs. {\bf 40}, 131, (1917)), questioned Slipher's results and Slipher replied (Obs., {\bf 40}, 304, (1917)), about the accuracy of the measurements and on the fact that the mean velocity of the spirals was found to be 570km/sec while the mean velocity of stars is about 20km/sec. As for the measurements of the speed of the Great Andromeda nebula Slipher got -300km/sec and he cited Wright who got -304km/sec, Wolf who got -300 - -400km/sec and Pease who got -329km/sec.}}. Campbell and Kapteyn discovered that stars with 'advanced' stellar spectra move faster.   Consequently, Campbell hypothesized that the radial velocities are associated with the evolutionary stages of the nebulae.  
Slipher explanation of the large velocities was that the nebulae are 'islands universe', namely stellar systems like ours at large distances. In other words, the spiral nebulae are extragalactic. And this was in 1917 several years before the Great Debate that took place in 1920.

\section{~First attempts to verify the de Sitter model}

In 1918, the couple  Shapley and Shapley{\footnote{Shapley, H.,  \& Shapley, M.B.,  ApJ., {\bf 50}, 107, (1919)}} analyzed  the differences in properties between globular clusters and spiral nebulae. The couple found that the radial velocities of globular clusters are predominately negative, a fact which  led  the  Shapleys  to the hypothesis  that  globular clusters are extra-galactic objects  and are falling onto the general galactic system.

As for the spiral nebulae, {\it the brighter spiral nebula as a class, apparently regardless of the gravitational attraction of the galactic system, are receding from the sun and from the galactic plane - a remarkable condition that has been little emphasized heretofore. }  The reason for this remarkable conclusion at this point in time was that though the number of velocities of spiral nebula was limited, out of 25 spirals all but 3 had positive velocities (meaning that they are moving away from us) and the velocities were $150 ~~km/sec$ and above. These velocities are much higher  (in absolute value) than the (negative) velocities of the  globular clusters.

The Shapleys stressed several times in the paper that {\it Globular clusters as a class appear to be rapidly approaching the galactic system; spiral nebula as a class are receding with high velocities.} They explained the fact that the spirals have receding velocities due to  repulsive forces which act between the spirals and the Milky Way.

Another conclusion of the Shapleys was that: {\it The speed of spiral nebula is dependent to some extent upon apparent brightness, indicating a relation of speed to distance or, possibly to mass. } This of course,  fitted the Shapley's idea of a repulsive force acting on the spirals, as they stated that: {\it The hypothesis demands that gravitation be the ruling power of stars and star clusters, and that a repulsive force, radiation pressure or an equivalent, predominate in the resultant behavior of spiral nebula.}

 It is interesting to note that the 1917 de Sitter's  paper was published just  a year before the Shapleys completed their work but in a quite obscure location{\footnote{de Sitter, W., Koninklijke Nederlandsche Akademie van Wetenschappen Proceedings, {\bf  19}, 367, 527, 1217, (1917)}}, not frequently accessed by astronomers. Neither were  the year earlier papers in the Monthly Notices  mentioned by Shapley (though his papers contained other references from this journal published roughly at the same time). On the other hand, the {\it Annalen der k.k 
 Universit${\ddot {\rm a}}$ts-Sternwarte in Wien } was consulted. The Shapleys had the first indication that  the system of spiral nebula expands. They did not know what the recession  velocity depended on, but the overall expansion was obvious and they stated that expansion may depend on distance.

After de Sitter's paper became  known various attempts to verify de Sitter's model were carried out. One of the first attempts  to discover the spectral shift (=velocity) versus distance relation was carried out by   Ludwig Silberstein (1872-1948){\footnote{Silberstein, L., MNRAS, {\bf 4}, 363, (1924)}} who  applied  de Sitter's result to derive a formula for the shift of the spectral lines emitted by stars.  For distant stars Silberstein found that in the limit of small velocities the Doppler shift is:
\begin{equation}
{\Delta \lambda \over \lambda}=\pm {r \over R}, \nonumber
\label{fig:silber}
\end{equation}
where $R$ is the radius of curvature of the universe and $\Delta \lambda$  the shift in wavelength $\lambda$ of the line.  Note that the sign can be positive or negative. Silberstein applied the formula to a list of stellar clusters for which he found mostly negative velocities but one positive one, as well as to the small and large Magellan Clouds which presented positive velocities. At that time it was not known that the Magellan clouds are outside the Milky Way. 
The application of this law to this mix of objects  gave a value for the radius of curvature of the universe of $R=6.0 \times 10^{12}$ astronomical units or about $10^8\:lyrs.$ In table \ref{tab:silber-1924} we give the data Silberstein used to derive $'$the radius of the universe$'$. All the objects in the table are globular clusters, save the Magellanic clouds.
As can be seen for example,  the implied distance to the  globular cluster   NGC 6934, is $6.7 \times 10^9 ~astronomical ~ units$ or  $100,000\:lyrs$ 
 or about the size of the Milky Way. It also implied that the small and large Magellan Clouds are outside the Milky Way. Silberstein sent his paper for publication on January 18th, 1924. The time  was before Hubble's discovery that the nebulae are extra-galactic. But, it appears that Silberstein was unaware of the astronomical debate because his results, though very inaccurate,  would have implied the same result Hubble got shortly after, namely that the spiral nebulae are extra-galactic.   On the other hand, the possibility that the globular clusters were as far away as he claimed, would have implied unacceptable high intrinsic luminosities for the stars in the cluster. In short, lack of astronomical facts gave rise to unacceptable results.

Eddington{\footnote{Eddington, A.S., Nature, {\bf 113}, 746, (1924)}} attacked Silberstein  proposal for determining the distance of a remote star by observing the displacement of the spectral lines at six months interval. Silberstein claimed that this method would  separate the ordinary Doppler effect of the unknown motion of the star from the distance-effect predicted by de Sitter. Eddington argued that  Silberstein neglected any effects of the short interval of time and of space between the two observations and of the distortion of the waves by the local gravitational field. 

On the other hand Eddington suggested the The ÒDouble DriftÓ Theory of Star Motions theory. The theory speaks about two observed stellar systems which differ in their mean velocity. The difference in mean speed, argued Eddington, must have arised in their formation because {\it it is difficult to see how gravitation towards the centre of the universe could separate the motions of the stars into two systems, if they originally formed one system}. 
\begin{figure}[t]
\centering
\epsfig{file=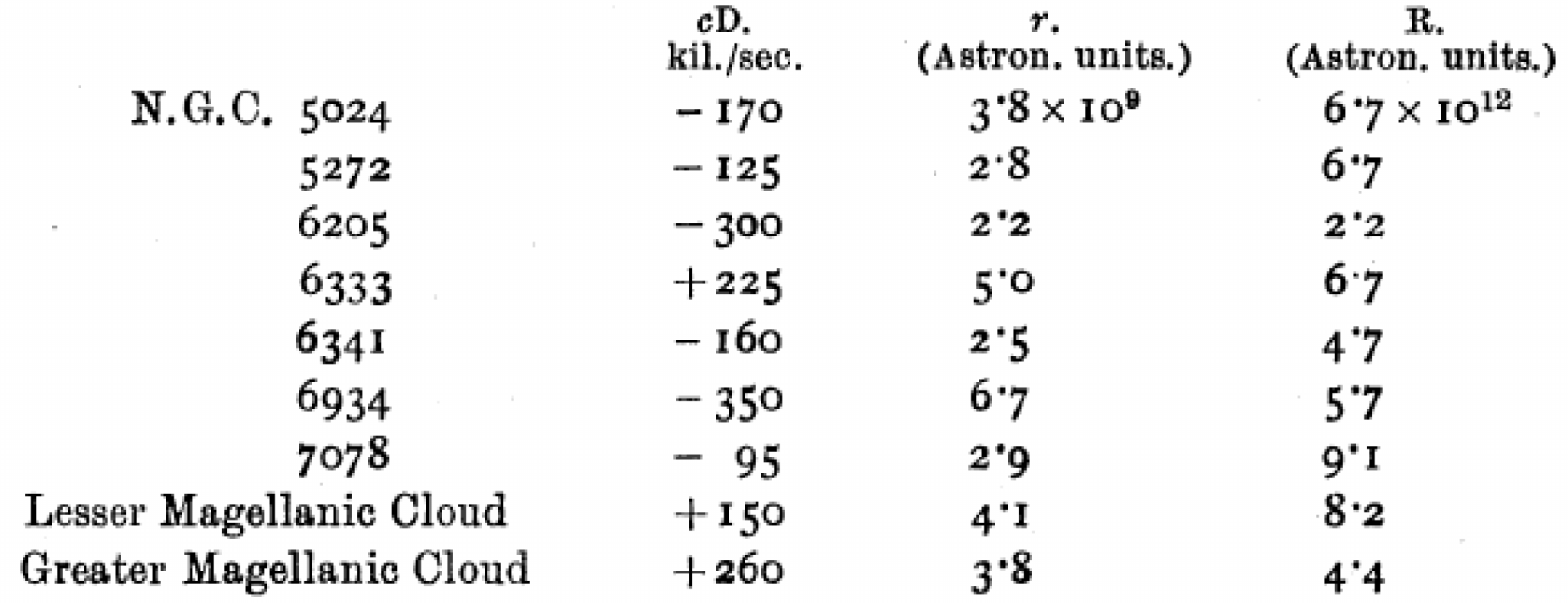,width=3.2in}
\caption{\sl  {\small The original data of Silberstein 1924.} } 
\label{tab:silber-1924}
\end{figure}

 Silberstein's results  were also criticized  by Lundmark(1889-1958){\footnote{Lundmark,K., MNRAS, {\bf 84}, 747, (1924) }}. The criticism reflects misfortune and how much  people can be locked up in dogmas. It is in this year that Hubble published his discovery of Cepheids in Andromeda and settled the "Island Universe" debate. Lundmark was still captive of the idea that the nebula were galactic. First, Lundmark claimed that the data on the Doppler shifts  which were taken from Slipher, were suspiciously large and  he questioned whether the  shifts in spectral lines were not caused by another reason which had nothing to do with velocity.  Lundmark could not believe that such high velocities can be real. Next, Lundmark pointed out that the theory of Weyl{\footnote{Weyl, H., Phys. Zeits., {\bf 24}, 230, (1923)}} and Eddington did not allow for a negative velocity, in contrast to Silberstein's result.  As for the globular clusters used by Silberstein, Lundmark claimed that: {\it These objects are probably among the most distant celestial objects we know at present, but how do we know that they are so far away that the effect of the curvature on space-time outweighs the effect of the real motions of the clusters themselves?} Lundmark repeated Silberstein's analysis for various groups of celestial objects. We show in fig. \ref{fig:lundmark-0} and
 \ref{fig:Lundmark+Hub} the data Lundmark compiled for the globular clusters and the spiral nebulae.  From the globular clusters data, Lundmark found a much greater radius of curvature for the universe than found by Silberstein. 
 
 Lundmark gave  a table of distances and velocities of spiral nebula. He based his distances on a {\it hypothetical parallax} derived from total magnitude and apparent diameter under the assumption that {\it the apparent angular dimensions  and the total magnitude of the spiral nebula are only dependent on the distance} which means they are standard candles.  Lundmark claimed that {\it An inspection of the table }(of distances and velocities){\it  will show that a computation of $R$ }(the radius of curvature of the universe) {\it from the individual values of V }(the velocity){it  will give inconsistent values for the radius of curvature.} However, no details were given. Next, Lundmark calculated that: {\it Plotting the radial velocities against these relative distances, we find that there may be a relation between the two quantities, although not very definite one. If this phenomenon were due to the curvature of space-time, we could derive the mean linear distance or determine the scale of our relative distance.}
 
 We took Lundmark's data for the 43 spiral nebulae  at face value, ignored the data for the Globular Clusters and assumed the existence of a linear relation, and calculated the Hubble constant to be $71.2\pm 55.6km/sec~ per~ Mpc$, a values extremely close to present day accepted value(!) but with a large uncertainty.   Lundmark, on the other hand, obtained   much larger numbers and consequently argued that the {\it simplified formula is not justified in this case}. Lundmark erred by lumping the globular clusters and the spiral nebula though he admitted that this lumping {\it certainly is open to objection.}  Lundmark's method gave much larger distances for the nebulae and consequently smaller Hubble constant. Evidently, Lundmark had a blind date with destiny and missed it.

 Shortly after Lundmark's criticism appeared in press, a similar criticism, this time by 
 Str${\ddot{\rm o}}$mberg, was published{\footnote {Str${\ddot{\rm o}}$mberg, G., ApJ., {\bf 61}, 353, (1925)}}. Str${\ddot{\rm o}}$mberg concluded that {\it we have found no sufficient reason to believe that there exists any dependence of radial motion upon distance}.

\begin{figure}[h]
\centering
\epsfig{file=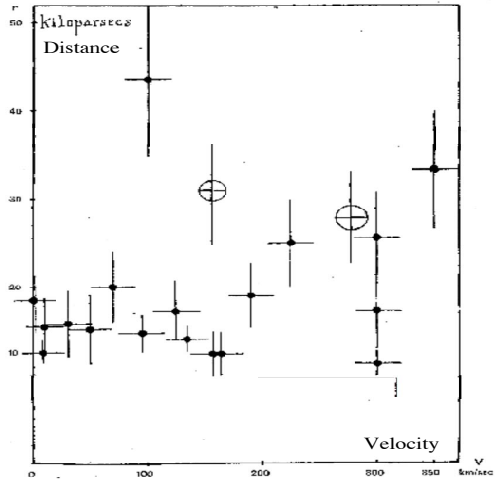,height=2.6in,width=2.6in}
\caption{\sl  {\small The relation between velocities and distances of globular clusters. The circles are the values for the two Magellan clouds. Lundmark, 1924. The signs of a linear correlations are seen?} } 
\label{fig:lundmark-0}
\end{figure}

Further attempts were carried out in 1922-1924 when Wirtz (1875-1939){\footnote{Wirtz, C., AN, {\bf 215}, 349, (1922)}} examined the statistics of the radial motions of spiral nebula. Out of 29 nebulae he had the data for, he found 25 with positive velocities so that the average speed of the spiral nebulae was $+840~~km/sec$. In 1924, Wirtz{\footnote{Wirtz, C., AN, {\bf 222}, 21, (1924)}} tried to 
relate  the velocities to the distance. To this goal, Wirtz  searched for a reliable distance indicator, for example, the apparent diameter, or luminosity of the spiral nebulae, assuming of course that they are all identical and can serve as $'$standard candles$'$.  Wirtz took the data of the apparent diameter of spiral nebulae from Curtis{\footnote{Curtis, H.D., Lick Public. \#13, (1918)}} and Pease,{\footnote{Pease, F.G., Mt. Wilson Cont. , {\bf 132},  (1917), {\bf 186}, 1920.}} to find out  that the velocity of the spiral nebula $v$ relates to the diameter $D$ as:
$$v( km/sec)=914-479 \cdot log( D).$$ Here the diameter is given in arc minutes. Wirtz was careful and added to this result also an evaluation of the reliability of the correlation just found.    In 1923 Eddington  derived $v \sim R^2$, a year later  Weyl obtained that $v \sim tan R$ and we mentioned before the Silberstein result. Hence, the logarithm was not a surprise. We can say that Wirtz almost confirmed de Sitter's model (we say almost because the expansion law in the de Sitter model does not contain the logarithm).

 Wirtz's contribution is hardly cited and when cited, it is by few historians of science. 
 In 1936, a short time before his death, Wirtz wrote a half a page long 
note{\footnote{Wirtz, C. Zeit. f. Astrophys., {\bf 11}, 261, (1936). {\it Ein literarischer Hinweis Zur Radialbewegung der Spiral Nebel}.}} reminding the reader about his discoveries back in 1921 and 1924, and before Hubble in 1929, but in no avail. 

\begin{figure}[h]
\centering
\epsfig{file=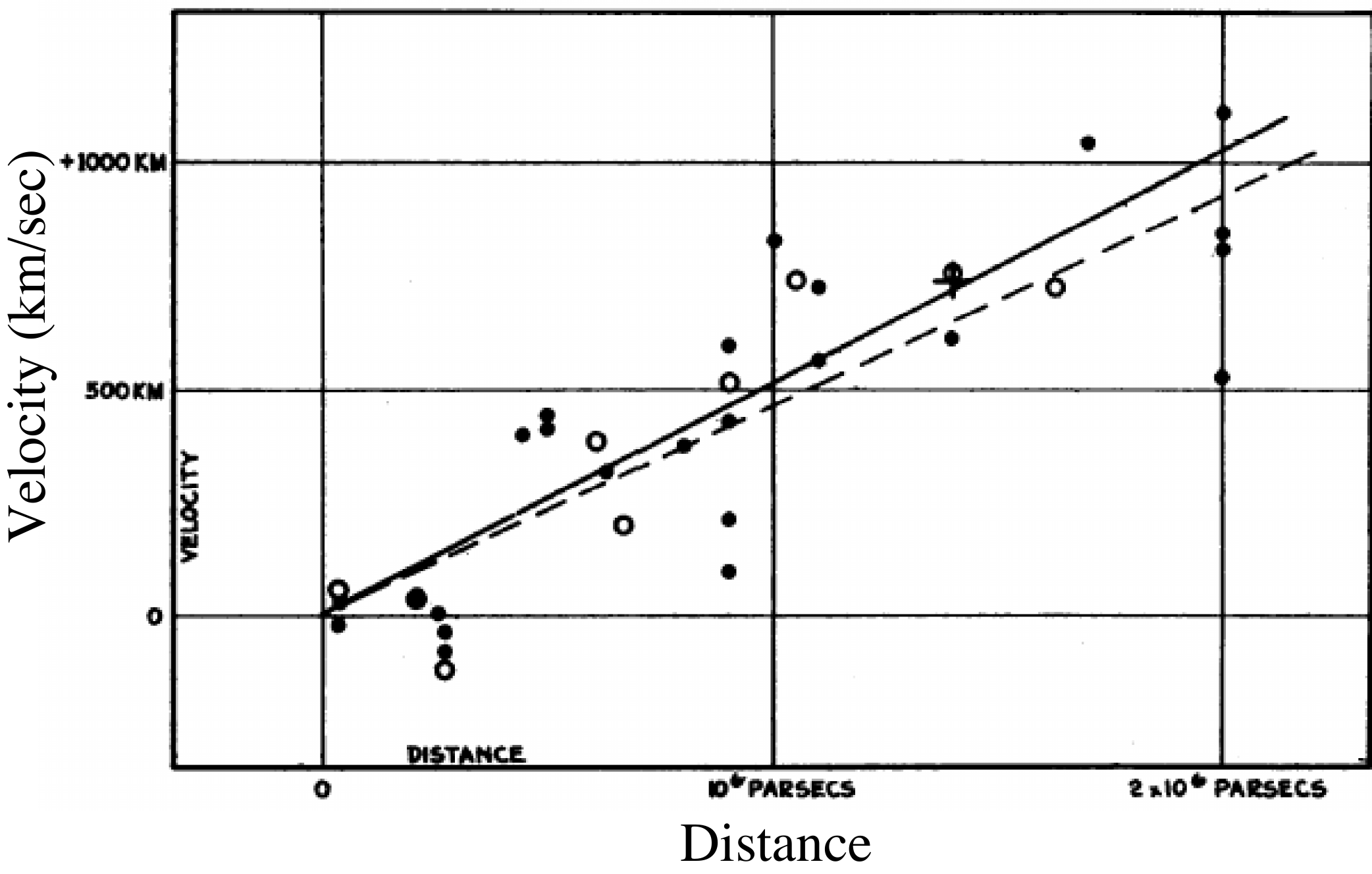,width=3.5in}
\caption{\sl  {\small The original velocity-distance relation discovered by Hubble in 1929.} } 
\label{fig:hubble-law}
\end{figure}
 
 While the measurement of the red-shift is difficult but straight forward, this is not the case with the distances. In 1926 Hubble carried out an extensive research on the distance of extra-galactic nebula and measured the distances of about 400 objects{\footnote{The objects were taken from a compilation by Hardcastle{\footnote{Hardcastle, J.A., MNRAS, {\bf 74}, 699, (1914)}} which was prepared from two hours exposure in a 10-inch telescope!}}. Hubble deduced a distance-apparent luminosity relation which improved the accuracy of the distances so measured. Also, further away nebulae were measured.

In 1929 
   Hubble (1898-1953){\footnote {Hubble, E., PNAS, {\bf 15}, 168, (1929). Published March 15.}}  decided to resolve the paradox that the {\it determination of the motion of the sun with respect to the extragalactic nebulae have involved a K term of several hundred kilometers which appears to be variable. Explanation of this paradox have been sought in a correlation between apparent radial velocities and distances, but so far the results have not been convincing. The present paper is a re-examination of the question, based on only those nebular distances which are believed to be fairly reliable.} Hubble had 24 pieces of data. According to Hubble, the data   indicated a linear correlation between distance and velocities. It is crucial to stress that Hubble did not assume a linear relation but plotted the data and demonstrated that a linear relation is a good approximation. We checked the numbers taking into account only the data from the table Hubble, assumed the existence of a linear correlation  and found for the coefficient of proportionality: $v=(470\pm278)r$. Hubble's original correlation is shown in fig.\ \ref{fig:hubble-law} and the proportionality coefficient in $500km/sec~per~Mpc$. In other words, Hubble succeeded to reduce the scatter in the data and obtain a convincing correlation and showed that there exists a correlation and it is linear. 
   
\begin{figure}[t]
\centering
\epsfig{file=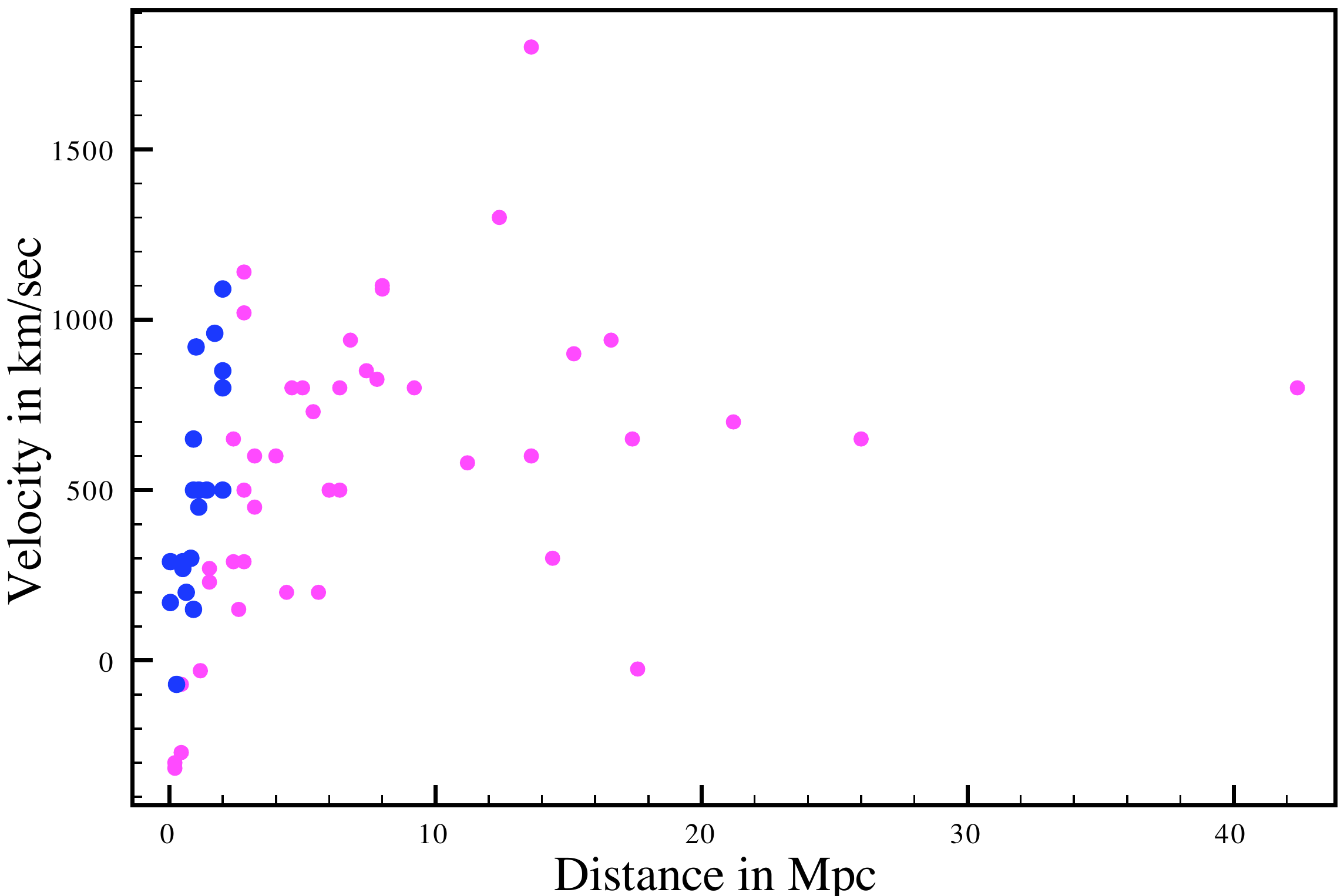,width=3.2in}
\caption{\sl  {\small The red data is from Lundmark (1925) 43 spiral nebulae while the blue one is from Hubble (1929) 23 spiral nebulae.} } 
\label{fig:Lundmark+Hub}
\end{figure}

In fig.\ref{fig:Lundmark+Hub} we compare the data used by Hubble and Lundmark. Hubble distances were significantly smaller than those of Lundmark and consequently the 'Hubble constant' constant that much larger.  
   
   In fig. \ref{fig:hubble-law1}    we can see what happened to the particular data used by Hubble. The red squares are the data of Hubble where distances were measured by Hubble and velocities by Slipher. The green squares are the additional data provided by Humason. The velocities of three nebulae originally measured by Slipher, are shown in blue. The arrow marks the   updated/changed velocities. One can see that the revision of the velocities improved significantly the correlation. Additional data, which is not shown in the figure, was provided by Humason, who measured a velocity of $3779km/sec$ for NGC 7619 at a distance of 7.8Mpc. 
   
\begin{figure}[h]
\centering
\epsfig{file=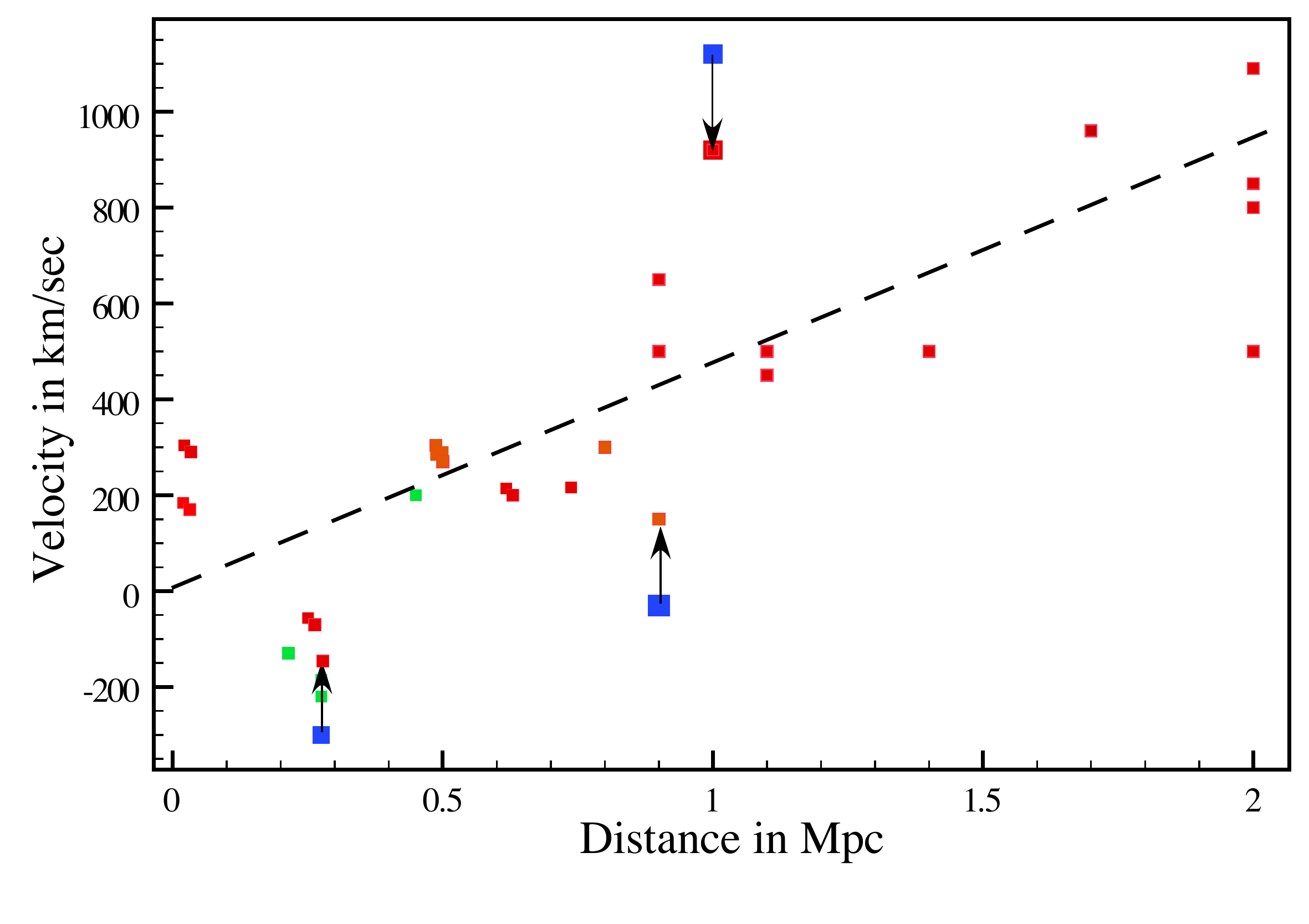,width=3.5in}
\caption{\sl  {\small The changes in the data used by Hubble in his velocity-distance correlation, 1929.} } 
\label{fig:hubble-law1}
\end{figure}

To appreciate Hubble's results we mention that  in October 1929 Perrine published{\footnote{Perrine, C. D., AN, {\bf 236}, 329, (1929). Submitted April 17, Published October 1929.}} an analysis of the motion of spiral nebulae and globular clusters, still mixing the two a couple of years after establishing the spirals as extragalactic objects. Perrine examined various correlations between the radial velocities and size of ellipticity, upon size, elongation and galactic latitude, upon distance, diameter and residual velocities and concluded that (A) {\it The radial velocities of the spiral and globular (structureless) nebula vary with apparent size, the smaller ones having the higher velocities.  (B) 	 The variation is not linear but increases rapidly among the smaller ones, being satisfactorily represented by the inverse square of the diameter}. So far so good, but he continued and concluded that:
 [\it (C) The relationship is to size (or mass) and not to distance.} There were additional conclusions all wrong probably because of poor statistics.
  Perrine did not assume that all spirals have the same size and  managed to show that the observations imply that the K-term is given by $K=V\sqrt{d}$, where $d$ is the apparent diameter.

\section{Criticism from astronomers}

Oort{\footnote{Oort, J.H., BAN, {\bf 5}, 105, (1931), Ibid. {\bf 6}, 155, (1931)}} criticized Hubble  claiming that: {\it It appears impossible to derive even a rough estimate of the distance of a cluster of nebulae as long as there may be such large and still undeterminable difference between the average luminosity of isolated and cluster nebulae as indicated in the first section. We may be surprised at the accuracy with which Hubble and Humason appear to be able to derive the relative distance of various clusters from apparent magnitudes, agreeing beautifully with the distance derived from radial velocities, but these do not help us in getting absolute values.} Hubble and Humason responded to the criticism without spelling out the criticizer, who was a renown astronomer,  by claiming strangely that it is of minor importance. {\it Differences tend to cancel out, and the final numerical result agrees with the Mt. Wilson result.} But it does not agree, as Oort's result (for the Hubble constant) was about half the result of Hubble and Humason.  

On the other hand, Hubble and Humason were very confident in their analysis and estimated, in their 1931 paper,   that {\it this number is correct to within $20\%$}. But its estimates  changed since then by a factor of 7!{\footnote{See: Trimble, V.,  $H_0$: {\it The Incredible Shrinking Constant 1925-1975}, PASP, {\bf  108}, 1073, (1996), how biases and improvements reduced substantially the Hubble constant.}}

\section{Understanding the recessional velocities}
The correlation found by Hubble was sufficiently accurate to cause people believe in it. At a meeting  of the Royal Astronomical Society early in 1930, de Sitter admitted that neither his nor Einstein's solution  could represent the observed universe. Eddington was stunned and asked {\it one puzzling question. Why should there be only these two solutions?} to Einstein equations. Eddington supposed that it was a consequence of the fact that people had only looked for static solutions. The participants of the meeting were unaware of Lemaitre's solution. Indeed, Lemaitre's 1927 paper was published in a practically   inaccessible journal not visited by astronomers. Consequently, it did not attract any attention and could rest in peace, if Lemaitre had missed the commotion in London. So Lemaitre wrote to Eddington, his previous mentor, and drew his attention to the discoveries made three years earlier. 

 Eddington{\footnote{Eddington, A.S., MNRAS, {\bf 90}, 668, (1930)}} recognized immediately the importance  of Lemaitre's paper and that his solution is not the right one. Eddington's paper is actually a review paper on Lemaitre's article  because {\it my original hope of contributing some definitely new results has been forestalled by Lemaitre's brilliant solution.} Eddington presentsed therefore, Lemaitre's solution giving him the full credit for the discovery that {\it the Einstein world is unstable}. Eddington praised Lemaitre solution and showed that Tolman's suggestion is unacceptable. However, when Eddington gave the recession velocity of the nebulae he just provided the number and he did not specify who obtained it let alone how he obtained it. Eddington noted the conflict between stellar ages and the age of the universe but did not offer any solution. 
 
 Eddington was convinced that Lemaitre's paper must be salvaged from its anonymity  and asked the paper to be translated into english and published in the MNRAS. In the meantime Lemaitre published in a journal much more accessible to astrophysicists{\footnote{Lemaitre, G., BAN, {\bf 5}, 273, (1930)}} a note explaining an apparent paradoxical  behavior of a mass point in an expanding universe. Lemaitre described de Sitter's solution but not his! The english version of the 1927 paper came out in 1931{\footnote{Lemaitre, G., MNRAS, {\bf 91}, 483, (1931), Ibid., 490, (1931). The only information in the paper is that it was translated by permission. No date is given and it is not clear whether Lemaitre saw the translation before publication. The translator of Lemaitre 1927 paper, as it appeared in the MNRAS,  is not specified. In the official site of the George Lemaitre center for Climate Research in Louvain (tttp://www.uclouvain.be/en-316446.html.) it is claimed that Eddington translated the article and added a long comment. I could not find any confirmation to this claim. }} in an abridged version. However, the  paper included the formula:
 $${R' \over R}={v \over cr}$$ 
 and the sentence: {\it From the discussion of available data, we adopt ${R' / R}=0.68\times 10^{-27}cm^{-1}$}. Here $R'/R=(R_2-R_1)/R_1$ where $R_1$ and $R_2$ are the radii at times $t_2$ and $t_1$ respectively. The idea is that this is a universal number. In Lemaitre's words, $v/c=(R_2/R_1)-1$ {\it it is the excess   radius of the universe when the light was emitted to the radius when the light is received.}  This figure amounts to $629km/sec~per~Mpc$. However, 
 no details as to how the number was calculated (see the details previously). Lemaitre noted that the numerical results imply: {\it quite recently for stellar evolution} but did not infer that there might be a conflict.

In  1930 Lemaitre  discussed the expanding universe{\footnote{Lemaitre, G. BAN. {\bf 5}, 273, (1930)}} and in 
 1933{\footnote{Lemaitre, G., PNAS, {\bf 20}, 12, (1934)}} he discussed the evolution of the expanding universe from an initial state. This was a revolution in thinking. The universe did not exists an infinite time but was created at some moment in time. This sounded to many like creationism and in particular when it came from a catholic priest! Consequently, many resented the idea. Lemaitre spoke about the mean density of clusters of nebulae and its connection to the {\it Hubble's ratio of distance to spectroscopic velocities by the approximate relation}
$${dr \over dt}=r \sqrt{{4\pi \over 3}G\rho_0}$$ and there is no mention of his own observational result for the coefficient of correlation. Lemaitre discussed  mainly the mean density $\rho_0$ and tried to infer the cosmic model from this parameter.

Naturally, the  discovery of the recessional velocities, even if only apparent,  stirred  discussions about the initial state out of which the expansion started. The questions were: (a) what causes the expansion and (b) what is the initial state which started the expansion.

One of the first explanations as to why the universe expands was given by Tolman{\footnote {Tolman, R.C., PNAS, {\bf  16}, 320, (1930)}} who suggested that the expansion is due to  the conversion of matter into radiation.  As the radiation accumulates the  radiation pressure increases and causes the expansion. In all these papers the physics of the initial state was not discussed and remained, therefore, an enigma. This explanation of the expansion was also suggested by Lemaitre{\footnote{Lemaitre, G., MNRAS, {\bf 91}, 483, (1931)}}.

  Lemaitre  provided  a possible explanation for the expansion of the universe: {\it  it seems to suggest that the expansion has been set up by  the radiation itself,} (cf. Tolman's explanation). In a static universe, light emitted by matter travels around the space, comes back to its starting point, and accumulates indefinitely. This light gives rise to radiation pressure which causes the universe to expand. 
 {\it seems that this may be the origin of the velocity of expansion which Einstein assumed to be zero, and which in our interpretation is observed as the radial velocity  of extragalactic nebulae,} explained Lemaitre.
 
In the second model published together with the first one{\footnote {Lemaitre, G., MNRAS, {\bf 91}, 490, (1931). The two models are published in the same journal issue, back to back.  }}, Lemaitre proposed that the universe started from an initial $'$stagnation$'$ state  which had extreme conditions. No further specification was given, in particular, neither the composition nor the temperature and density were specified. 

     \section{~Did Hubble believe in an expanding universe?}
  It seems that   at no time did Hubble believed that the spectral wavelengths  shifts represent real velocities. At the end of the correlation discovery paper, after establishing that the K-term depends linearly only on distance, he pointed out that {\it The outstanding feature, however, is the possibility that the velocity-distance relation may represent the de Sitter effect} \dots {\it the displacement of the spectra arise from two sources, an apparent slowing down of atomic vibrations and a general tendency of material particle to scatter} the words expansion and universe were not mentioned at all. As a matter of fact, Hubble remained skeptical of 'The expansion of the Universe' and refrained from using such a term, even when the rest of the community, and in particular the theoreticians, accepted the idea of cosmic expansion. 
  
  Objectively,  Hubble had good reasons at that time to suspect this explanation. His determination of the 'Hubble constant' yielded an age of $2\times 10^9$ years for the universe, while Jeans estimated that the stars are about a factor $10^3$ older 'and stellar structure was more advanced'. 
  
  More specifically, Hubble had several technical objections. In 1936{\footnote{Hubble, E. , ApJ, {\bf 84}, 517, (1936)}} Hubble thought that: 
 {\it It is evident that the observed result, (of applying a  K-correction B=2.94,) is accounted for if the redshifts are not velocity shifts.} and in the discussion of the paper Hubble expressed his doubts in the expanding universe as a possible solution. Additional, mainly technical points were raised by Hubble over the years{\footnote{Hubble, E. , {\it The Observational Approach to Cosmology}, Oxford: Clarendon Press, 1937.}}.

 At the end of their paper, Hubble and Humason iterated that: {\it The interpretation of red-shift as actual velocities, however, does not command the same confidence, and the term 'velocity' will be used for the present in the sense of 'apparent' velocity, without prejudice as to its ultimate significance. }  The authors do not provide an error estimate because of possible systematic errors in the magnitudes and color indices however, they state that: {\it It is believed, however, that the uncertainty in the final result is definitely less than $20\%$.}  They trace the difference in this result to {\it  the revision Shapley made in the standard unit of distance. }

   In 1935 Hubble and Tolman{\footnote{Hubble, E. \& Tolman, R. C., ApJ, {\bf 82}, 302, (1935	)}} wrote that: {\it The most obvious explanation of this finding is to regard it as directly correlated with a recessional motion of the nebulae}\dots {\it Nevertheless, the possibility that the red-shift may be due to some other cause } \dots {\it should not prematurely neglected} \dots {\it Until further evidence is available, both the present writers wish to express an open mind with respect to the ultimately most satisfactory explanation.}
   
   In 1942 Hubble wrote in an article to science{\footnote{Hubble, E., Science, {\bf 95}, 212, (1942)}}  that; {\it It may be stated with confidence that red shifts either are velocity shifts or they must be referred to some hitherto unrecognized principle in nature.}  Moreover, after an analysis Hubble concluded that: {\it Thus the assumption that red-shifts are not velocity shifts but represent some hitherto unknown principle operating in space between the nebulae leads to a very simple, consistent picture of a universe so vast that the observable region must be regarded as an insignificant sample.}

  In 1953 Hubble delivered the George Darwin lecture{\footnote{Hubble, E.P., MNRAS, {\bf 113}, 658, (1953). Hubble passed away before he was able to revise the paper before publication, as he intended. The paper was edited by A. R. Sandage.}} and discussed {\it The law of red-shifts}. Hubble noted three phases in the history, the first of which was {\it a discovery phase which ended with a crude formulation in 1928-29}. There was no mention of Lemaitre. Hubble argued that {\it if the red-shifts do measure the expansion of the universe} then we can trace its history. In the paper, Hubble rejected the term 'apparent velocity' and used just 'velocity' but meant $c d\lambda/\lambda$ or 'red shifts expressed on a scale of velocities'. More specifically, Hubble raised again the conflict between the age of the Earth and the implied age for the Universe, which emerges if the expansion solution is assumed. And more technically {\it When no recession factors are included, the law will represent approximately a linear relation between redshift and distance. When recession factors are included, the distance relation (becomes) nonlinear.  }
     
Few months after Hubble published his sensational correlation, Zwicky{\footnote{Zwicky, F., PNAS, {\bf 15}, 773, (1929), Phys.Rev., {\bf 34}, 1623, (1929), Ibid. {\bf 48}, 802, (1935),}}, also at Caltech, found another explanation for the red-shift, namely gravitational drag of light. This suggestion explains why the frequency shift is wavelength independent. Zwicky was even capable of providing a theoretical estimate to the Hubble constant given the mean density in the Universe. 

In recent years Burbidge et al{\footnote{Burbidge, G., Hoyle, F., \& Schneider, P. A\&A, {\bf 320}, 8, (1997)}} and Arp{\footnote{Arp, H.
{\it Current issues in cosmology}. Eds. Pecker \& Narlikar. Cambridge University Press,  2006.}}
have consistently discovered  very close pairs of Quasi-Stellar-Objects with separations $<5 arcsec$  but very different redshifts. The  estimates that these situations are due to  random projection are very small. Consequently, they conclude either that this is  evidence that QSOs have significant non-cosmological redshift components, or that the pairs must be explained by gravitational lensing.

 \section{Could the omission be intentional?}
 
 Some historians of science expressed the view that reference to previous works was not a common practice in those days. I reject this claim. First, Hubble himself cited de Sitter simply because he believed in his model. Second, there are  papers in those days which contain many references. Third, we are in the opinion that Hubble did not cite Lemaitre   deliberately because he did not believe in his model. 
 
 The claim that Hubble overlooked the paper by Lemaitre while he cited de Sitter ignores the situation. Hubble published his paper in 1929 and  Lemaitre published his work in 1927  in the   Annals de la Soci$\acute{\rm e}$t$\acute{\rm e}$ Scientifique de Bruxelles. How many  astrophysical and astronomical libraries carry today this publication? Let alone in the past. How many astronomers check today for relevant papers in this and similar publications?   What do you want from Hubble, who was in California, when Eddington, who was on the other side of the English channel,  stated that he overlooked Lemaitre's paper{\footnote{Eddington, A.S., MNRAS, {\bf 90}, 668, (1930)}}.  de Sitter who was in neighboring Leiden was unaware 
 of Lemaitre's publication. 
 
 In 1930, de Sitter published in America{\footnote{de Sitter, W., PNAS,{\bf  16}, 474, (1930)}} a discussion about the expansion of the Universe. de Sitter was in Holland, the neighboring country to Belgium, where Lemaitre published, and so wrote de Sitter: {\it A dynamical solution of the equations(4),with the line-element (5), (7) and the material energy tensor (6) is given by Dr. G. Lemaitre in a paper published in1927, which had unfortunately escaped my notice
until my attention was called to it by Professor Eddington a few weeks
ago.}   de Sitter does not mention any  'Lemaitre's determination of the Hubble constant'.

During the 1927 Solvay meeting  Lemaitre discussed his solution with Einstein. Lemaitre spoke already on an expanding universe as a result of two forces, gravity (potential of $1/R$) and repulsive (potential of $1/R^2$). Einstein was not convinced at that time and continued to hold the view that the universe is static.
In 1930 Einstein published a paper{\footnote{Einstein, A.  Zum kosmologischen Problem der allgemeinen RelativitŠtstheorie, in Albert Einstein: Akademie-VortrŠge: Sitzungsberichte der Preu§ischen Akademie der Wissenschaften 1914-1932 (ed D. Simon), Wiley-VCH Verlag GmbH \& Co. KGaA, Weinheim, FRG, (2006)}} and admited an expanding universe but failed to cite or mention Lemaitre's contribution.

  Few american scientists read then and today european journals. When Ritter published in 1878-1883 in German his seminal series of papers of the theory of stellar structure,  Hale, the then editor of the Astrophysical Journal, complained in a special editorial introduction to Ritter's paper, that {\it These seems to have received hardly the attention they deserve, and at the suggestion of one of our associates we have translated one of the most important papers on the series.}{\footnote{Ritter, A. ApJ, 8, 293, (1898)}} As a matter of fact, the paper was re-edited and abridged by the editorial board of the ApJ and the name of the translator is not given. 

 \section{Why Lemaitre did not complain?}
 In 1950 Couderc{\footnote{Couderc, P., {\it L'expansion de l'Univers},  Presses Universitaires de France, Paris, 1950.}} published a book about the expanding universe. 
 Lemaitre{\footnote{Lemaitre, G., Ann. Ap., 13, 344, (1950)}} did not like few of the descriptions in the book and  wrote to the editor of the then French professional journal Les Annales d'Astrophysique. 
 
 Lemaitre's letter provides us with a rare   glimpse into his personality. Lemaitre admited to be embarrassed when he discussed the last two chapters  in the book in which his work is favorably exposed.  What bothered Lemaitre was (a) the Couderc moved to the appendix other theories which were heterodox and (b) The "legendary" description of the chain of events. 

Next, Lemaitre pointed that he does not want to argue if he is a professional astronomer (as described by Couderc) though he was a member of the IAU and studied astronomy for two years (one under Eddington and one "in an american observatory"). Lemaitre mentioned that he visited Hubble and Slipher and he heard Hubble announcing in 1925 his memorable discovery of the distance to Andromeda.  As for the expansion of the universe, Lemaitre wrote that {\it while my mathematical bibliography is seriously faulty, it is perfect from astronomical point of view. I calculated the coefficient of expansion (575km/sec per Mpc or 625 with a doubtful statistical correction). }Then Lemaitre added an interesting sentence and I bring it in french: "Naturellement, avant la d$\acute{\rm e}$couverte et l'$\acute{\rm e}$tude des amas de n$\acute{\rm e}$buleuses, il ne pouvait 
$\hat{\rm e}$tre question d'$\acute{\rm e}$tablir la loi de Hubble"  or, naturally, before the discovery of and the study of clusters of nebula, there was no question of establishing the Hubble law but only calculating the coefficient. According to Lemaitre, the title of his note does not leave any doubt about his intentions: A Universe with constant mass and increasing radius explains the radial velocities of the extra-galactic nebula. He then continues to discuss the instability which formed the nebula from a super-dense state. Lemaitre ends his note by raising the possibility that the Universe decreased and then rebounds like a phenix. 
Lemaitre, as he himself wrote, assumed a linear relation, then calculated the mean velocities and mean distances and divided. 

In 1952, Lemaitre discussed the "Clusters of Nebulae in an Expanding Universe"{\footnote{Lemaitre, G., MNSSA, {\bf 11}, 110, (1952).}} and wrote that: {\it Hubble and Humason established from observation the linear relation between distance and velocity which was expected for theoretical reasons and which is known as the Hubble velocity-distance relation. Hubble's ratio between a distance of one mega-parsec and the velocity there found of 55 km/sec, amounts to 2 thousands million years. It constitute the main feature and fixes the scale of what is called the "expanding universe"}. Lemaitre does not mention that he measured  the number previously and obtained a close number. Moreover, Lemaitre argued that the large dispersion in the velocities cannot be overlooked in any discussion of the velocity-distance relation and must be carefully considered before accepting the reality of some observed departure from it.

\section{The 'censored' paper}
Recently, the insinuation that Hubble, or someone on his behalf,  censored the translation of Lemaitre's paper into english was tossed into the air{\footnote{For a report on the claim see: Reich, E.S., Nature,  {\it Edwin Hubble in translation trouble}, 27 June, 2011.}}. The implication is that in this way Lemaitre was deprived of the credit as a discovered of what erroneously is called the Hubble law.
The idea that Lemaitre's 1931 translation was censored is not raised by  the official site of the George Lemaitre center on its web page. Moreover, as we have shown, the number obtained for the 'Hubble constant' is given in the translation, though the method how it was derived is not specified. Also, we stress the difference between assuming a linear relation as Lemaitre did and proving the existence of a linear relation, as Hubble did.  

The translator of Lemaitre 1927 paper, as it appeared in the MNRAS,  is not specified. In the official site of the George Lemaitre center for Climate Research in Louvain{\footnote{http://www.uc;ouvain.be/en-316446.html.}} it is claimed that Eddington, who was Lemaitre former superviser in his master studies in Cambridge, translated the article and added a long comment. If this is so, we suppose that Eddington made the distinction between what Lemaitre assumed and Hubble proved.

In 1930 de Sitter published three papers{\footnote{de Sitter,  W., BAN, {\bf 5} 157,   211, 274, (1930), PNAS, {\bf  16}, 474, (1930)}}. In the first paper de Sitter discussed extragalactic nebulae and their distances and apparent radial velocities and concluded that: {\it The only acceptable explanation of these large receding velocities so far proposed is by the inherent expanding tendency of space which follows from the solution, that was called solution (B) in my paper of 1917, of Einstein's differential equation for the inertial field. It will be remembered that these equations admit two, and only two solutions which are static and isotropic.}  de Sitter mentioned Lemaitre's non-static solution from 1927, which he admited, has escaped his notice and to which his attention was called by Eddington. de Sitter did not mention in this paper Lemaitre's result for the correlation between velocity and distance. 
De Sitter's next paper{\footnote{de Sitter, W., BAN, {\bf 5}, 211, (1930)}} was devoted to a discussion of Lemaitre's solution and repeated Lemaitre solution which implies an expanding universe. Also here there is no mention of Lemaitre's determination of the proportionality coefficient. 
In  a third paper by de Sitter{\footnote{de Sitter, W., BAN, {\bf 5}, 274, (1930)}}  he further  discussed astronomical consequences of the theory of the expanding universe without mentioning  Hubble's or Lemaitre's determination of the constant.
Last, in 1930 de Sitter  discussed Lemaitre's solution and even showed his  (de Sitter's) figure which proves the distance-radial velocity correlation, but no mention that  the numerical coefficient was found by either Hubble nor Lemaitre. 

In 1952 Lemaitre discussed clusters of nebula in an expanding universe{\footnote{G. Lemaitre, MNSSA, {\bf 11}, 110, (1952)}}. Lemaitre distinguished carefully between nebulae and clusters of nebulae and related the velocity-distance relation to the clusters and not to the individual nebulae. He considered the individual nebulae in equilibrium in the clusters and the motion in the cluster as the peculiar motion on top of what we call today the Hubble flow. 
As for our point here, Lemaitre discussed what he called the {\it Hubble velocity-distance relation} and did not mention that he calculated the coefficient already in 1927.

Eddington in his book {\it The Expanding Universe} did not mention Lemaitre's derivation of the Hubble constant.
And so wrote Eddington:{\it The simple proportionality of speed to distance was first found by Hubble in 1929.} The law is predicted by relativity (de Sitter). Furthermore: {\it According to the original investigation of de Sitter a velocity proportional to the square of the distance would have been expected; but the theory had become better understood since then, an it was already known (though perhaps only to a few) that simple proportionality to the distance was the correct theoretical result.} Eddington comments that he was not among the few who knew about it in 1929.

\section{Summary of the history of numerical determination}
In table \ref{tab:early-H} we give chronological order and the values of the early measurements of the Hubble constant. 
\begin{table}[htdp]
\caption{Early measurements of the 'Hubble constant'}
\begin{center}
\begin{tabular}{lllll}
Author & year & value & Number & comments\cr
           &          &          & of objects  &   \cr
           Wirtz & 1922-24 & 840 &\cr
Lundmark & 1925 & 71 & 43 & Calculated on the basis of his data \cr
Lemaitre & 1927 & 625  & 42 & Weighted \cr
Lemaitre & 1927 & 670  & 42 & unweighted \cr
Hubble & 1929 & 530 & 24 \cr
de Sitter & 1930 & 460 & 54 \cr
Eddington & 1930 & 522 &\cr
Lundmark & 1930 &   604  & 25 &Spiral-formed PASP, {\bf 42}, 23, 38, (1930)\cr
 Lundmark   & 1930   & 876& 17 &Elliptical  PASP, {\bf 42}, 23, 38, (1930)  \cr
Oort & 1931 & 290 & 44 \cr
Hubble \& Humason & 1931 & 558 & 68 & Special treatment for clusters of galaxies \cr
\end{tabular}
\end{center}
\label{tab:early-H}
\end{table}%

\section{So why  is it called the Hubble law?}
Eddington, 1931{\footnote{Eddington, A.S., MNRAS, {\bf 92}, 3, (1931). Also Proc. Phys. Soc. {\bf 44}, 1, (1932)}} described the recession of the galaxies, credited Friedman and Lemaitre with providing the theory but wrote that: {\it Hubble showed further that the speed of recession is approximately proportional to the distance.} 

As far as we could check, Hubble never cited Lemaitre even when he discussed the possible implications of the red-shift-distance correlation years later{\footnote{Hubble, E., MNRAS, {\bf 113}, 658, (1953)}}. Very plausibly  because he simply did not believe in an expanding universe. Let alone, Lemaitre's expanding universe model removed from oblivion by Eddignton in 1930 and  was non-existent  for Hubble in 1929. 

The paper by Hubble and Humason 1931 is preceded by a paper by Humason with the title {\it Apparent velocity-shifts in the spectra of faint nebulae}. Humason starts with: {\it In 1929 Hubble found a relation connecting the velocities and distances \dots} No mention of any theory of theoretician. Similarly, Humason and Hubble 1931 do not mention Lemaitre at all.

Tolman from Caltech{\footnote{Tolman, R.C. ApJ, {\bf 69}, 245, (1929)}}, before Lemaitre's work was exposed on the pages of the MNRAS,   was probably the first to credit Hubble, who was also from Caltech, with the discovery. Tolman wrote that {\it The correlation between distance and apparent radial velocity for the extra-galactic nebulae obtained by Hubble}. Tolman then states that {\it Consideration has already been give to these relations by de Sitter, Weyl, Eddington and other, and in particular by Silberstein, who obtained conclusions not previously appreciated.} This was probably the origin of the credit to Hubble. The date on Tolman's paper in the Astrophysical Journal is 25 February, 1929, while the date on Hubble's paper is 15 March, 1929. Thus Tolman knew about the paper prior to publication. Tolman and Hubble wrote joint papers and their offices were both in Pasadena.  Zwicky was in Pasadena at the time Hubble published his results and in his first attempt in 1929 to explain the observations he wrote that: {\it Hubble has shown recently that the correlation between the apparent velocity of recession and the distance is roughly linear}.

Oort in 1931 claims that the correlation is due to the work on mount Wilson and later mentions Hubble and Humason's contribution. Slipher and Flagstaff are not mentioned. 

In 1933 Eddington published his book: {\it The Expanding Universe}{\footnote{Eddington, A.S., {\it The Expanding Universe}, Cambridge press, 1933}} and on 
  page 46 you find his version: {\it The deliberate investigation of non-static solution was carried out by Friedmann in 1922. His solutions were rediscovered in 1927 by Abbe Lemaitre, who brilliantly developed the astronomical theory resulting therefrom. His work was published in a rather inaccessible journal, and seems to have remained unknown until 1930 when attention was called to it by de Sitter and myself.  In the meantime the solution was discovered the third time by Robertson, and through his their interest was beginning to be realized. The astronomical application , stimulated by Hubble and Humason's observational work on the spiral nebula, was also being rediscovered, but it had not been carried so far as in Lemaitre's paper. }
  
  In 1933 Robertson wrote a long review{\footnote{Robertson, H.P., RMP, {\bf 5}, 62, (1933)}} about the expanding universe. He cited Hubble, Hubble and Humason and even Silberstein, but no mention of Lemaitre who obtained a solution with mass and calculated the coefficient between the velocity and the distance. However, Lemaitre paper from 1931 about the expanding universe, was cited.

  In 1939 McVittie{\footnote{McVittie, G.C., Proc. Phys. Soc. {\bf 51}, 529, (1939)}} wrote a review. The review does not mention Lemaitre's theoretical nor observational work.  

Sandage, A. 1972 starts his paper with: {\it Hubble (1929) discovered the velocity-distance relation by correlating red-shift and apparent magnitude of nearby galaxies. This correlation had been previously suggested by Robertson (1928) but was not discussed by him}. No  mention of Lemaitre or de Sitter. Lemaitre yok!

Last, Lemaitre in 1929, before Hubble's publication, did not cite his number for the coefficient of proportionality.

 \section{To whom should the credit go?}
  Different people select different researchers as the predictors of the expansion of the universe. However,  de Sitter, Friedman, Lemaitre, Robertson and Tolman found dynamic expanding solutions and hence all qualify as predictors of the expansion of the universe.  What about Einstein whose equations are at the center of the problem and  did not believe at first   in a dynamic universe and change his mind?
  
  As for the observers, 
Kragh and Smith{\footnote{Kragh, H. \& Smith, R.W., His. Sci. {\bf 41}, 141, (2003)}} argue that {\it Hubble cannot reasonably be credited with the discovery of the expanding universe.} Strictly speaking, Hubble himself did not claim it. Actually, their paper discusses the problematics of the entire notion of the  discovery of the expanding universe. It is not the first time that the credit should be divided as well as the name of the phenomenon. We do not enter this philosophical discussion but would like to give two other examples.    The equation for ionization in stellar atmospheres was discovered by Eggert{\footnote{Eggert, J., Phys. Zeit., 20, 570, (1919)}}. Saha{\footnote{Saha, M.N., Phil. Mag., 40, 472, (1920)}} just fixed the constant to be the ionization energy. Now it is called the Saha equation. Eddington discovered the Kramers opacity law{\footnote{Eddington, A.S. MNRAS, 83, 32, (1922)}} and Kramers{\footnote{Kramers, H.A., Phil. Mag., 46, 836, (1923)}} fixed the constant in the law discovered by Eddington. Now it is called Kramers opacity. (see Shaviv{\footnote{Shaviv, G., {\it The Life of Stars, The Controversial Inception and Emergence of the Theory of Stellar Structure}, Springer \& Magnes press, 2009.}}).  Should Lundmark, Lemaitre and Hubble be equally credited? What about Slipher?

As for the provocative title of this paper, Hubble who found the data used by others had no reason to plagiarize and did better and more accurately than others. He proved for the first time  what other tried to prove or assumed. 

{\bf Acknowledgement}
E-mail exchanges with Sidney van den Bergh are highly appreciated, though some of our views are not alike. 
\end{document}